\documentclass{article}
\usepackage{spconf,amsmath,graphicx,hyperref}
\usepackage{siunitx}
\usepackage{amsmath}
\usepackage{amssymb}
\usepackage{xcolor}
\usepackage{enumitem}
\usepackage{flushend}
\usepackage{enumitem}
\def\MM{\text{DECODE}}

\title{DECODE: Dual-Enhanced COnditioned Diffusion for EEG Forecasting}
\name{Mehran Shabanpour$^{\dagger}$, Sadaf Khademi$^{\dagger}$, Konstantinos N Plataniotis$^{\ddagger}$, and Arash Mohammadi$^{\dagger}$
\thanks{This work was partially supported by Natural Sciences \& Eng. Research Council (NSERC) of Canada; NSERC Discovery Grant RGPIN-2023-05654.}}
\address{$^{\dagger}$Concordia Institute for Information Systems Engineering (CIISE), Concordia University, Canada.\\
$^{\ddagger}$ Department of Electrical and Computer Engineering, University of Toronto, Canada}
\begin{document}
\ninept
\maketitle
\begin{abstract}
Forecasting Electroncephalography (EEG) signals during cognitive events remains a fundamental challenge in neuroscience and Brain-Computer Interfaces (BCIs), as existing methods struggle to capture both the stochastic nature of neural dynamics and the semantic context of behavioral tasks. We present the Dual-Enhanced COnditioned Diffusion (DECODE) for EEG, a novel framework that unifies semantic guidance from natural language descriptions with temporal dynamics from historical signals to generate event-specific neural responses. DECODE leverages pre-trained language models to condition the diffusion process on rich textual descriptions of cognitive events, while maintaining temporal coherence through history-based Langevin dynamics. Evaluated on a real-world driving task dataset with five distinct behaviors, DECODE achieves sub-microvolt prediction accuracy (MAE~$=~0.626~\mu\text{V}$) over 75-timestep horizons while maintaining well-calibrated uncertainty estimates. Our framework demonstrates that natural language can effectively bridge high-level cognitive descriptions and low-level neural dynamics, opening new possibilities for zero-shot generalization to novel behaviors and interpretable BCIs. By generating physiologically plausible, event-specific EEG trajectories conditioned on semantic descriptions, DECODE establishes a new paradigm for understanding and predicting context-dependent neural activity.
\end{abstract}
%
\vspace{-.15in}
\section{Introduction}\label{sec:intro}
\vspace{-.1in}
\setlength{\textfloatsep}{0pt}
\begin{figure*}[t]
\centering
\includegraphics[scale=0.52]{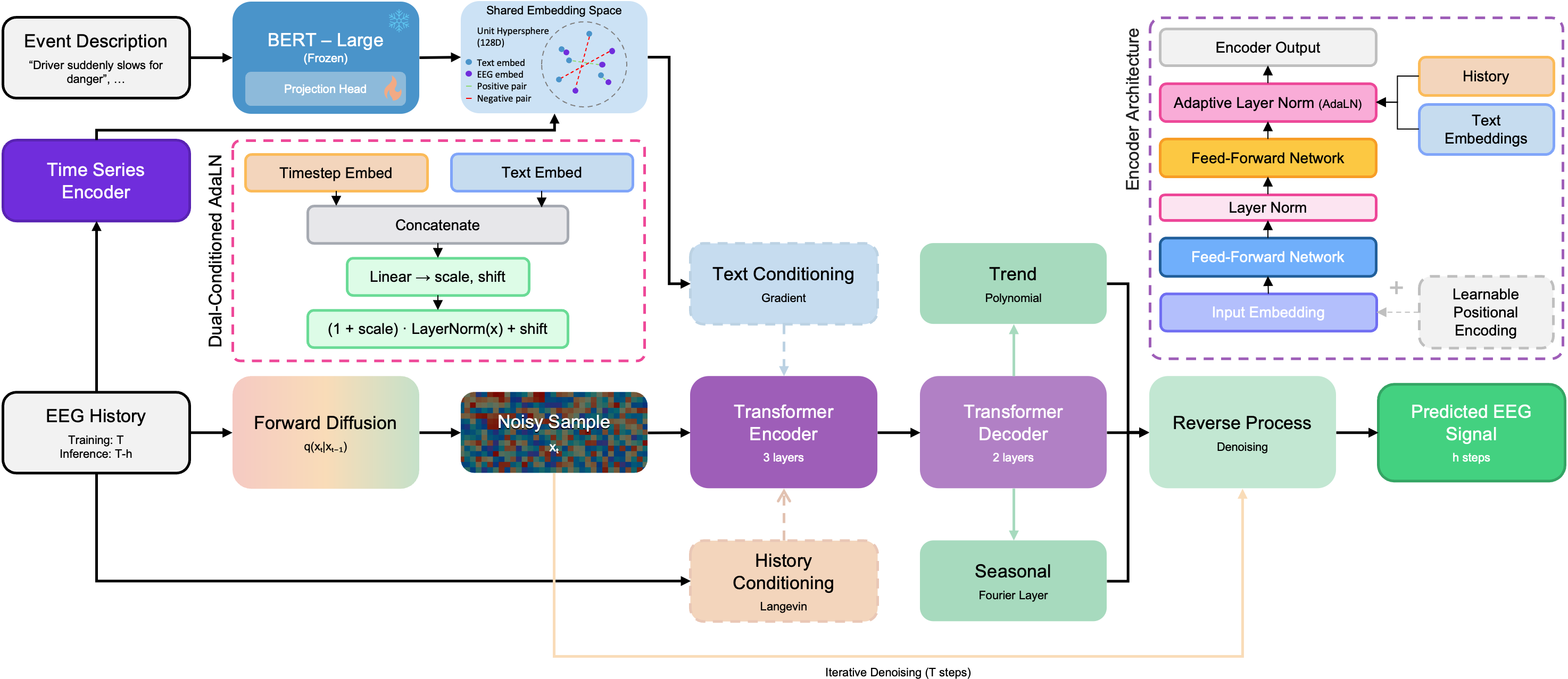}
\vspace{-.1in}
\caption{\footnotesize Architecture of the proposed dual-conditioned diffusion model for event-specific EEG forecasting. The DECODE framework integrates semantic guidance from natural language event descriptions (top path) with history-based temporal conditioning (bottom path) through a hierarchical diffusion process. Text descriptions are encoded via frozen BERT-large and projected to a shared embedding space, while historical EEG signals undergo forward diffusion before processing through encoder-decoder transformers with interpretable decomposition.}
    \label{fig:model-architecture}
\vspace{-.2in}
\end{figure*}
The ability to accurately forecast future brain patterns based on historical neural signals has become increasingly critical for applications ranging from seizure prediction to real-time cognitive load assessment and brain-state-dependent stimulation protocols~\cite{Jiang2024}. In this context, Electroencephalography (EEG) signals provide a window into the complex dynamics of brain activity, offering unprecedented opportunities for understanding cognitive states, predicting neural responses, and developing Brain-Computer Interfaces (BCIs). The inherently non-stationary, multi-scale, and subject-specific nature of EEG signals, however, presents fundamental challenges that have limited the effectiveness of traditional forecasting approaches. The complexity of EEG forecasting extends beyond simple time series prediction. Neural signals exhibit intricate spatio-temporal correlations across multiple frequency bands~\cite{Chang2022, DiLiberto2021}, with dynamics that vary across individuals, cognitive states, and environmental~contexts. 

Recent studies have demonstrated that cognitive events and behavioral responses manifest as distinct patterns in EEG recordings, particularly in high-stakes scenarios such as driving tasks where split-second neural responses can determine critical outcomes \cite{Tao2024, Angkan2024}. These observations underscore the need for forecasting models that can capture both the stochastic nature of neural dynamics and the deterministic patterns associated with specific cognitive events.
Traditional approaches to EEG forecasting, including autoregressive models and Recurrent Neural Networks (RNNs), have shown limited success in capturing the full distribution of possible future trajectories, particularly when conditioned on complex cognitive events~\cite{Pankka2025}. 
While these methods can estimate conditional means effectively, they struggle to model the uncertainty and multi-modal nature of future EEG states, particularly when conditioned on complex cognitive events. This limitation becomes especially pronounced in real-world applications where understanding the range of possible neural responses is as important as predicting the most likely outcome.
The paper targets addressing this limitation by capitalizing on recent advances in generative modeling, particularly, diffusion models, which have revolutionized time series forecasting~\cite{Li2024, Yuan2024}. Unlike traditional discriminative approaches, diffusion models learn to reverse a gradual noising process, allowing them to capture the full conditional distribution of future states while maintaining temporal coherence \cite{Huang2025, Wang2025}. For the targeted task of EEG forecasting, this capability is particularly valuable as it enables the generation of physiologically plausible signals that respect the complex spectral and spatial characteristics of neural activity.

\noindent
\textbf{\textit{Literature Review:}} Diffusion models for time series forecasting have progressed from early conditional approaches such as TimeGrad~\cite{Li2024}, which suffered from error accumulation and limited long-range dependencies, to self-supervised masking strategies such as CSDI and SSSD, enabling parallel generation \cite{Ye2025, Yuan2024}. More recent works have focused on enhancing conditioning and representation; TimeDiff~\cite{Shen2023} addresses boundary issues, TMDM~\cite{Li2024} leverages transformer architectures capturing uncertainty in multivariate forecasting tasks, Diffusion-TS~\cite{Yuan2024} preserves trend and seasonality via Fourier-based losses, CN-Diff~\cite{Rishi2025} integrates nonlinear time-dependent transformations to capture complex temporal patterns, and D3U~\cite{Li2025} separates deterministic and uncertain components for improved probabilistic forecasting. Despite these advances, existing methods primarily rely on historical observations or simple categorical conditions, lacking the flexibility to incorporate rich semantic information about future events.

When it comes to EEG forecasting, traditional models relied on linear autoregressive models, which remain practical but perform poorly for long-range predictions. Recent deep learning methods, such as WaveNet~\cite{Pankka2025}, enhance conditional probability modeling of future samples, particularly for theta and alpha bands. Multi-channel EEG forecasting, despite its importance for capturing whole-brain dynamics, has received limited attention. EEGDIF~\cite{Jiang2024} addressed this gap with a diffusion model achieving correlation coefficients above $0.77$ across $16$ channels. However, most approaches remain limited to resting-state or seizure EEG, with little focus on event-conditioned forecasting in complex cognitive tasks.
Application of diffusion models to EEG data presents unique challenges due to the signals' complex spatio-temporal structure and subject-specific variations. EEGDfus~ \cite{Huang2025} introduced a dual-branch CNN-Transformer architecture for denoising, highlighting the importance of multi-scale feature extraction. DiffEEG~\cite{Shu2025} leveraged diffusion-based data augmentation conditioned on short-time Fourier spectrograms to address class imbalance in seizure prediction. Other approaches focus on EEG reconstruction and synthesis, with STAD~\cite{Wang2025} using multi-scale Transformers for super-resolution reconstruction, and classifier-free guidance generating subject-, session-, and class-specific Event-Related Potentials (ERPs)~\cite{klein2024}.
Understanding EEG dynamics during cognitive tasks provides crucial context for developing event-specific forecasting models. EEG studies in driving tasks have identified distinct neural signatures associated with different driving behaviors. In particular, beta-band Event-Related Desynchronization/Synchronization (ERD/ERS) and fronto-parietal activity reflect motor preparation and cognitive load during driving, highlighting their relevance for modeling task-specific neural dynamics into forecasting models~\cite{Tao2024, Angkan2024}.

\noindent
\textbf{\textit{Contributions:}} We propose a novel conditional diffusion framework for event-specific EEG forecasting that incorporates semantic information to guide signal generation, enabling preemptive safety interventions in semi-autonomous vehicles. Our key insight is that natural language descriptions of events contain rich contextual information that can guide the diffusion process toward generating event-appropriate neural responses. This intuition goes beyond discrete event labels, capturing the richness of cognitive events~\cite{Tao2024}. In summary, the paper makes the following contributions:
\begin{itemize}[leftmargin=10pt, itemsep=0pt, topsep=0pt]
\item The Dual-Enhanced COnditioned Diffusion (DECODE) framework is introduced, which is the first diffusion-based EEG forecasting model that simultaneously leverages natural language descriptions for semantic guidance and historical signals for temporal coherence. The proposed dual conditioning mechanism addresses the critical challenge of latent space collapse when generating signals with subtle inter-class differences.
%
\item Introduction of a Semantic-Neural Bridge (SNB) within the underlying contrastive learning approach by aligning Bidirectional Encoder Representations from Transformers (BERT)-encoded text descriptions with EEG temporal patterns. The SNB provides zero-shot capabilities to handle novel behavioral conditions and continuous interpolation between cognitive states, unavailable in existing discrete-label approaches.
\end{itemize}
The proposed $\MM$ is evaluated over challenging real-world scenarios, achieving significant improvements in deterministic metrics with particular emphasis on event-specific prediction accuracy. More specifically, $\MM$ achieves sub-microvolt prediction accuracy of $0.626 \mu$V  while maintaining well-calibrated uncertainty estimates (CRPS $< 0.72$). It outperformed its closest comparable method by approximately $24$\% on multi-channel event-related forecasting tasks.

\vspace{-.1in}
\section{The $\MM$ Framework} \label{sec:framework}
\vspace{-.1in}
We consider the task of conditional EEG signal forecasting, where given a historical EEG sequence $\mathbf{x}_{1:T-h} \in \mathbb{R}^{(T-h) \times d}$ and a natural language description $c$ of an upcoming neural event, we aim to predict the future EEG signals $\mathbf{x}_{T-h+1:T} \in \mathbb{R}^{h \times d}$ that correspond to the described event. Here, $T$ denotes the total sequence length, $h$ represents the forecasting horizon, and $d$ is the number of EEG channels. 
Next, the utilized dataset will be briefly introduced followed by the detailed developments of the $\MM$ framework.

\vspace{.05in}
\noindent
\textbf{2.1. Dataset Overview}

\noindent
We utilize the Multimodal Physiological Dataset for Driving Behaviour (MPDB)~\cite{Tao2024}, which contains synchronized EEG recordings from 35 participants performing naturalistic driving tasks in a high-fidelity simulator. The dataset captures five distinct driving behaviors: braking, turning, lane changing, acceleration, and stable driving (baseline condition), with event markers precisely aligned to behavioral onsets.
For this study, we focus exclusively on the 59-channel EEG data sampled at 1000~Hz. Each trial spans 2 seconds (500~ms pre-stimulus to 1500~ms post-stimulus) centered on event markers. The raw dataset comprises approximately 5,700 trials across all participants and conditions.

\vspace{.05in}
\noindent
\textbf{2.2. Pre-processing and Data Preparation}

\noindent
EEG data from 30 participants across five driving tasks were preprocessed using artifact rejection, bandpass filtering (0.1–30 Hz), baseline correction, and average referencing. ERPs were extracted from 4,308 trials, and 14 electrodes most sensitive to driving events were selected by computing mean amplitudes within a 0-500 ms post-stimulus window across all conditions and ranked channels by their absolute deviation from baseline. For event-specific forecasting, the selected electrodes were Fpz, P7, P8, T8, PO7, O1, PO8, PO5, CP2, Pz, C2, CP3, CP4, and FC4, capturing both frontal positivity and central-parietal negativity. Time-frequency analysis was performed across theta, alpha, beta, and low gamma bands using Hilbert-based power estimation. ERD/ERS was computed as the percentage power change from a pre-stimulus baseline to an active window, and independent t-tests with Cohen's $d$ effect sizes identified band-condition pairs with the highest discriminative power for forecasting cognitive-motor states. 

\begin{figure}[t]
  \centering
  \includegraphics[scale=0.36]{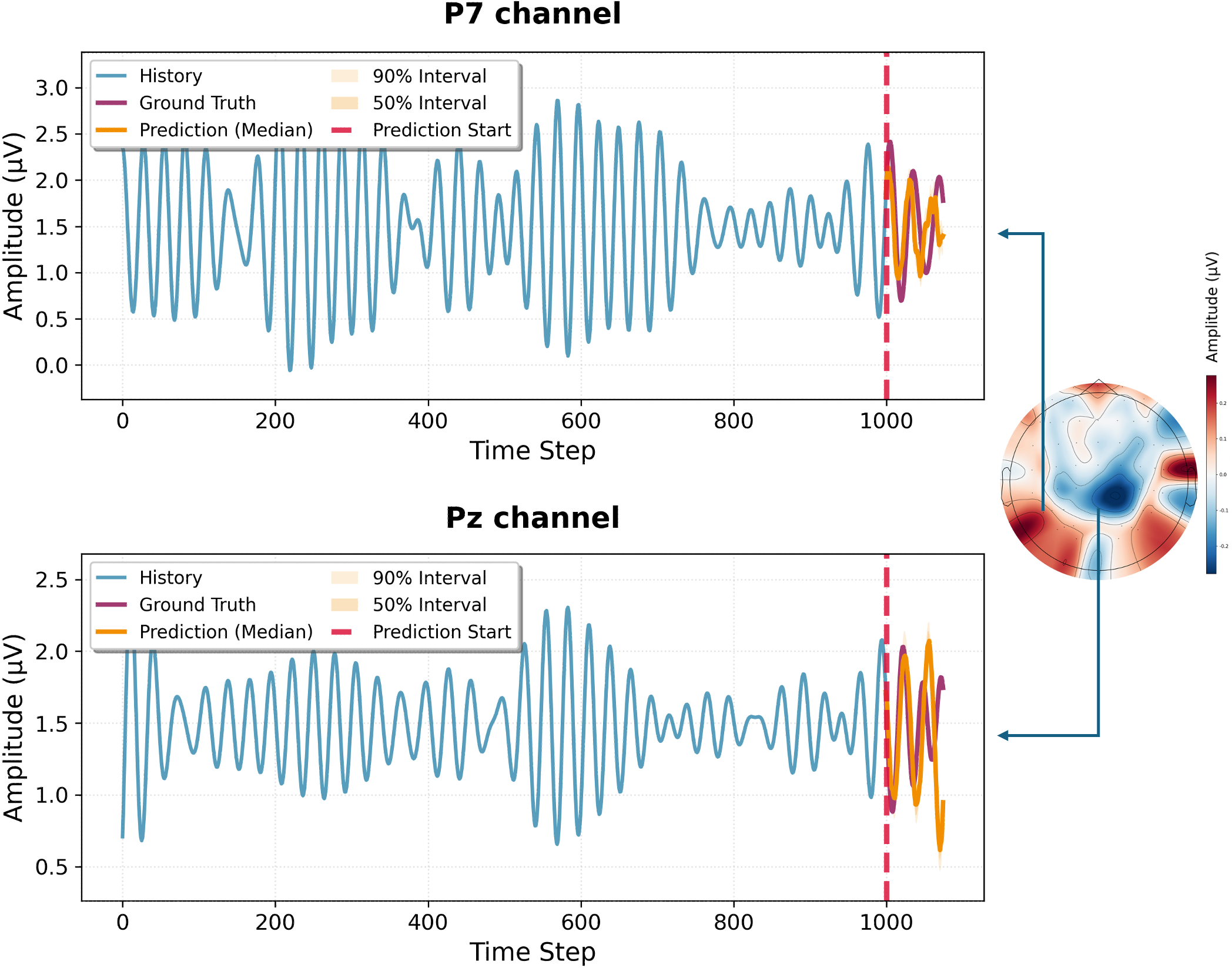}
\vspace{-.1in}
  \caption{\footnotesize Representative EEG forecasting results for electrodes P7 and Pz, with historical signal (blue) providing temporal context up to timestep 1000 and the model generating probabilistic forecasts (orange) compared to ground truth (purple). The topographic map shows grand-average voltage distribution (0-500 ms post-stimulus) with red indicating positive and blue indicating negative potentials.}
  \label{fig:erp_topo}
  \vspace{.1in}
\end{figure}

\vspace{.05in}
\noindent
\textbf{2.3. Base Diffusion Model}

\noindent
Our approach builds upon the Diffusion-TS architecture~\cite{Yuan2024}, employing a denoising diffusion probabilistic model with interpretable decomposition. The forward diffusion process gradually corrupts the clean signal \(\mathbf{x}_0\) through \(T\) timesteps according to 
\begin{equation}
    q(\mathbf{x}_t \mid \mathbf{x}_{t-1}) = \mathcal{N}\left(\mathbf{x}_t; \sqrt{1 - \beta_t} \, \mathbf{x}_{t-1}, \beta_t \mathbf{I} \right),
\end{equation}
where \(\beta_t\) follows a cosine schedule~\cite{nichol2021}.
The reverse process learns to denoise through a transformer-based architecture that explicitly models trend and seasonal components
\begin{equation}
 \hat{\mathbf{x}}_0(\mathbf{x}_t, t, \theta) = V_{tr}^t + \sum_{i=1}^{D} S_{i,t} + R,   
\end{equation}
where \(V_{tr}^t\) represents the trend component modeled via polynomial regression in a low-frequency space, \(S_{i,t}\) captures seasonal patterns through Fourier synthesis layers that select dominant frequencies via top-\(k\) amplitude selection, and \(R\) denotes the residual component~\cite{Yuan2024}.
The model employs an encoder-decoder transformer with adaptive layer normalization conditioned on the diffusion timestep, where each decoder block progressively refines the decomposition through cross-attention with encoded representations. Figure~\ref{fig:model-architecture} illustrates the complete architecture of our dual-conditioned diffusion framework.

\vspace{.05in}
\noindent
\textbf{2.4. Text-Guided Semantic Conditioning}

\noindent
To prevent latent space collapse and enable semantic control over generated EEG patterns, we introduce a text conditioning mechanism using pre-trained language models. We employ BERT-large~\cite{devlin2019} to encode natural language event descriptions \(\{c_k\}_{k=1}^K\) into a shared embedding space with time series representations. 
As shown in the top pathway of Figure~\ref{fig:model-architecture}, the text encoder \(f_{\text{text}}\) projects BERT's pooler output through a learned projection head
\begin{equation}
    \mathbf{e}_{\text{text}} = \text{Normalize}(\mathbf{W}_{\text{proj}} \cdot \text{BERT}(c) + \mathbf{b}_{\text{proj}}),
\end{equation}
where normalization ensures unit sphere embeddings. 
Correspondingly, the time series encoder \(f_{\text{ts}}\) maps EEG sequences through convolutional layers followed by projection to the same embedding space. We optimize a contrastive objective based on the InfoNCE loss~\cite{oord2018} 
\begin{equation}
    \mathcal{L}_{\text{contrast}} = -\log \frac{\exp(\mathbf{e}_{\text{ts}} \cdot \mathbf{e}_{\text{text}}^+ / \tau)}{\sum_{j} \exp(\mathbf{e}_{\text{ts}} \cdot \mathbf{e}_{\text{text}}^j / \tau)},
\end{equation}
where \(\tau\) is a learned temperature parameter and \(\mathbf{e}_{\text{text}}^+\) denotes the correct text embedding. 
This alignment creates semantically meaningful regions in the latent space, preventing mode collapse while maintaining behavioral coherence.

\vspace{.05in}
\noindent
\textbf{2.5. Dual Conditioning Mechanism}

\noindent
During inference, we employ a dual conditioning strategy that combines history-based Langevin dynamics with text-based gradient guidance. For history conditioning, we apply Langevin sampling to enforce consistency with observed signals \(\mathbf{x}_\text{obs}\) through iterative refinement
\begin{equation}
    \mathbf{x}_{t-1}' = \mathbf{x}_{t-1} + \eta_h \nabla_{\mathbf{x}_t} \left[ \alpha \|\mathbf{x}_\text{obs} - \hat{\mathbf{x}}_\text{obs}\|^2 + \gamma \log p(\mathbf{x}_{t-1} \mid \mathbf{x}_t) \right],
\end{equation} 
where \(\eta_h\) controls the step size, \(\alpha\) weights reconstruction fidelity, and \(\gamma\) balances generation fluency~\cite{song2021}. The number of Langevin steps \(K\) is adaptively scheduled based on the diffusion timestep, with more iterations at higher noise levels where structural decisions are made.
For text conditioning, we compute gradients of the log-probability with respect to the predicted class:
\begin{equation}
    \nabla_{\mathbf{x}_t} \log p(c \mid \mathbf{x}_t) = \nabla_{\mathbf{x}_t} \text{LogSoftmax}\left( f_\text{ts}(\mathbf{x}_t) \cdot \mathbf{E}_\text{text}^\top / \tau \right)[k],
\end{equation}
where \(\mathbf{E}_\text{text}\) contains pre-computed embeddings for all event descriptions and \(k\) indexes the target event.
The final update combines both guidance signals with tuned weights to ensure history dominates local dynamics while text provides semantic drift toward appropriate behavioral manifolds:
\begin{equation}
    \mathbf{x}_{t-1} = \mu_\theta(\mathbf{x}_t, t) + \sigma_t \mathbf{z} + \lambda_h \mathbf{g}_\text{history} + \lambda_t \mathbf{g}_\text{text}.
\end{equation}
This hierarchical conditioning creates a Riemannian metric on the latent manifold where geodesics follow natural behavioral transitions while maintaining semantic coherence.

\begin{figure}[t]
  \centering
  \includegraphics[scale=0.43]{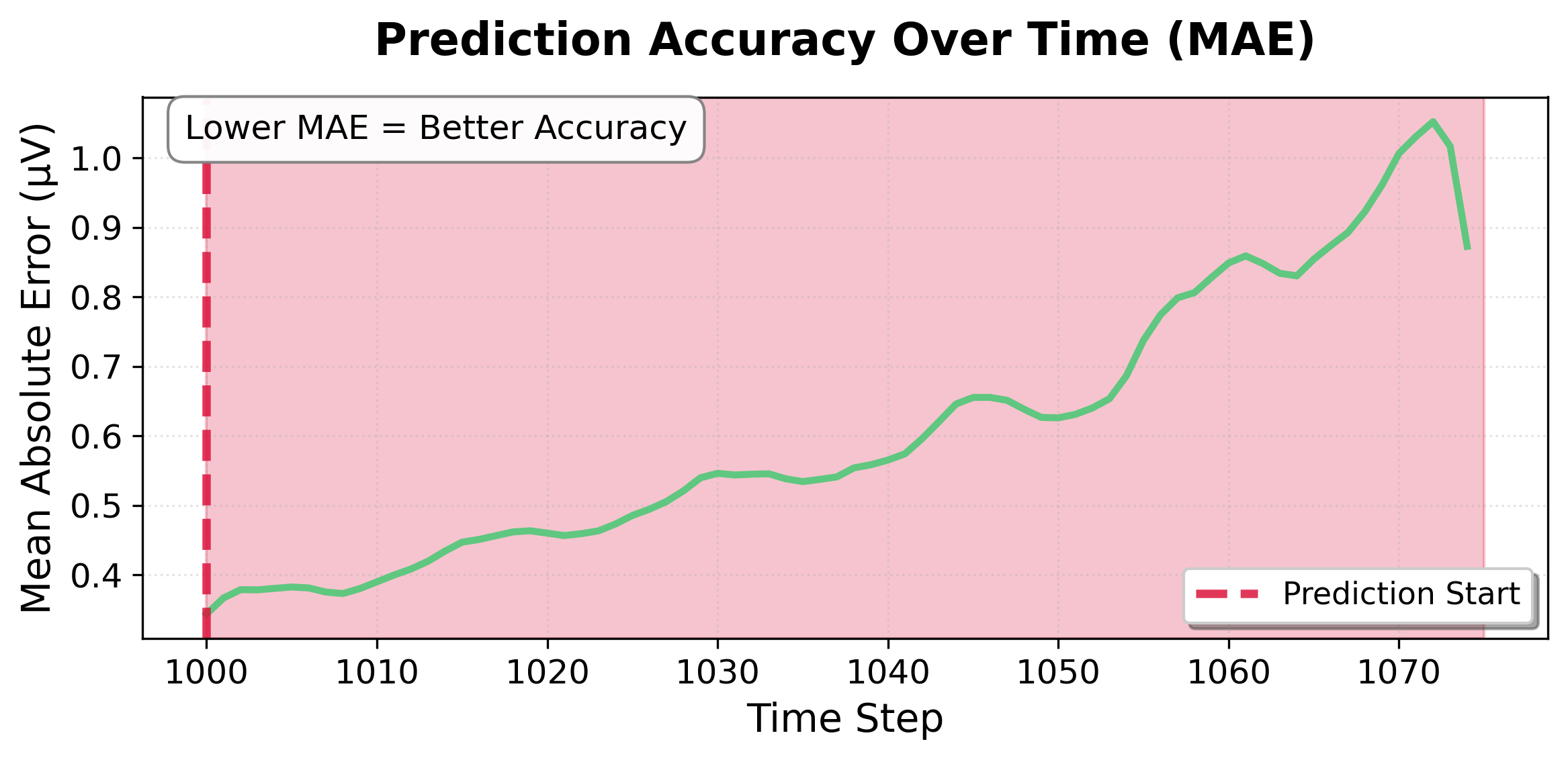}
  \vspace{-.1in}
  \caption{\footnotesize Mean Absolute Error (MAE) over time across the 75-timestep prediction horizon.}
  \label{fig:G3}
\end{figure}
\section{Experimental Results} \label{sec:results}
Our experiments employed a dual-conditioning diffusion model with a Transformer-based architecture consisting of 3 encoder layers, 2 decoder layers, hidden dimension of 96, and 4 attention heads for multivariate EEG forecasting. The diffusion process utilized 500 training and 200 sampling timesteps, with a cosine beta schedule and \(L_1\) loss optimization. Training proceeded for 12,000 epochs on sequences of length 1,075 with 59-dimensional features, using the Adam optimizer (learning rate \(= 1 \times 10^{-5}\)) and Exponential Moving Average (EMA) with decay rate 0.995 for stabilization. The dual conditioning mechanism balanced history-based Langevin dynamics with text-guided gradient conditioning using frozen BERT-large embeddings.

\vspace{.05pt}
\noindent
\textbf{\textit{Data Characteristics and Augmentation Strategy:}} EEG data presents unique challenges compared to conventional time series benchmarks. While datasets like \textit{Traffic} and \textit{Electricity}~\cite{Haoyi2020} represent continuous temporal processes with arbitrary segmentation preserving long-range dependencies, EEG consists of genuinely independent trials with no inter-trial causal relationships. Each trial represents a discrete experimental event with statistical independence, eliminating inter-trial temporal dependencies and forcing the model to learn generalizable patterns purely from within-trial dynamics. 
We addressed this through sliding window augmentation with 64-timestep strides, ensuring each window contained at least one event marker. This increased the effective training set size by approximately 15-fold while preserving temporal structure. The subtle differences between EEG patterns across driving events, evidenced by the moderate 83.59\% classification accuracy, present particular challenges for unconditional diffusion models, which are prone to mode collapse when class boundaries are ambiguous.

\vspace{.05in}
\noindent
\textbf{\textit{Neural Response Characterization:}} The oscillatory dynamics of individual electrodes were captured, and the grand-average topographic distributions (Fig.~\ref{fig:erp_topo}) revealed distinctive frontal-parietal dissociation patterns, with frontal positivity and parietal negativity modulations following an urgency-dependent hierarchy. Table~\ref{tab:neural_responses} presents the quantitative statistical analysis of neural responses across driving events. Brake events elicited the strongest cortical responses across both ERP and spectral domains, consistent with their higher cognitive demands for rapid motor inhibition and decision-making.
Notably, our spectral analysis revealed ERS rather than the classical desynchronization typically associated with motor tasks, suggesting enhanced cortical engagement during complex driving behaviors. Beta and low gamma bands emerged as the most discriminative features (Cohen's $d>0.18$), with synchronization magnitudes directly correlating with behavioral urgency. 
Accordingly, we selected the low-gamma band for subsequent analyses due to its strong discriminative power. This unexpected ERS pattern may reflect the heightened attentional demands and sensorimotor integration required for real-time vehicular control, distinguishing these naturalistic driving tasks from simpler motor paradigms.

\vspace{.05in}
\noindent
\textbf{\textit{Forecasting Evaluation:}} Our approach represents a novel paradigm in EEG forecasting, and differences in signal amplitude, prediction and history window length, and frequency bands, make direct comparison with other studies challenging. 
The most relevant baseline, Pankka et al.'s WaveNet approach \cite{Pankka2025}, achieved MAE values of \(1.0 \pm 1.1~\mu\text{V}\) for theta and \(0.9 \pm 1.1~\mu\text{V}\) for alpha bands over 150~ms (75 time steps) horizons on single-channel resting-state data. 
We further benchmarked TimeGrad \cite{Rasul2021} and PatchTST \cite{Nie2023}, two complementary state-of-the-art time series forecasting models, on our dataset, obtaining average MAEs of \(0.759~\mu\text{V}\) and \(0.733~\mu\text{V}\), respectively.
In comparison, our model achieves an MAE of \(0.626 \pm 0.29 ~\mu\text{V}\) for low gamma band at 75-timestep horizons on multi-channel event-related recordings with semantic conditioning. According to Fig.~\ref{fig:G3}, MAE exhibits characteristic degradation with increasing forecast horizon, demonstrating the inherent trade-off between temporal reach and prediction accuracy as the model operates under progressively weaker observational constraints.

\begin{table}[t]
\centering
\caption{\footnotesize Statistical analysis of neural response characteristics across driving events.}
\vspace{.1in}
\label{tab:neural_responses}
\small
\setlength{\tabcolsep}{4pt}
\begin{tabular}{llcccc}
\hline
\textbf{Measure} & \textbf{Event} & \textbf{Value} & \textbf{$p$} & \textbf{$d$} \\
\hline
\multicolumn{5}{l}{\textit{ERP Modulation ($\mu$V)}} \\
\hline
Frontal (Fz) & Brake & +1.06 & $<$.001 & -- \\
             & Turn & +0.81 & .015 & -- \\
             & Lane Change & +0.70 & .005 & -- \\
Parietal (Pz) & Brake & -1.42 & $<$.001 & -- \\
\hline
\multicolumn{5}{l}{\textit{Event-Related Synchronization (\%)}} \\
\hline
Beta (13-30 Hz) & Brake & +16.0 & .0002 & 0.18 \\
Gamma (30-45 Hz) & Brake & +27.7 & $<$.0001& 0.23 \\
\hline
\end{tabular}
\begin{minipage}{\linewidth}
\vspace{5pt}
\small
\textbf{Note:} p-values indicate significance levels as follows: p$<$0.05 (significant), p$<$0.01 (highly significant), and p$<$0.001 (very highly significant).
\end{minipage}
\end{table}

\vspace{.05in}
\noindent
\textbf{\textit{Ablation Analysis:}} We conduct ablation studies to identify the individual contribution of history conditioning and to determine how its integration with text guidance drives overall performance. History conditioning alone (MAE \(= 0.704~\mu\text{V}\)) provides temporal coherence but lacks event-specific guidance. The synergistic combination achieves superior performance (MAE \(= 0.626~\mu\text{V}\)), validating our architectural choices. 
The learned embeddings support continuous interpolation between behavioral conditions, suggesting potential for generating intermediate cognitive states not explicitly present in training data. This capability, combined with the model's ability to maintain semantic separation through text embeddings while preserving temporal coherence through history conditioning, addresses the fundamental challenge of generating context-dependent neural signals with subtle inter-class differences.

\vspace{-11pt}
\section{Conclusion}
\vspace{-9pt}
\label{sec:conclusion}

This work introduces DECODE, a dual-enhanced conditional diffusion framework that unifies semantic guidance and temporal dynamics for event-specific EEG forecasting. By combining natural language descriptions with history-based temporal conditioning, we enable generation of physiologically plausible and behaviorally coherent neural trajectories, achieving sub-microvolt prediction accuracy with well-calibrated uncertainty estimates.
The continuous nature of learned embeddings indicates potential for generating intermediate cognitive states through embedding interpolation, while the probabilistic formulation with calibrated uncertainty makes the approach suitable for safety-critical BCI applications.
The iterative denoising paradigm offers an unexplored advantage for real-time BCIs: early termination of the reverse process yields partially denoised signals that may preserve sufficient task-relevant information while dramatically reducing inference latency, suggesting future work should investigate optimal noise-fidelity trade-offs for time-critical neural interfaces.


\bibliographystyle{IEEEbib}
\footnotesize
\bibliography{strings,refs}

\end{document}